\documentclass[conference]{IEEEtran}

\input{00_Header}

\begin{document}

\title{Inferring Latent Market Forces: Evaluating LLM Detection of Gamma Exposure Patterns via Obfuscation Testing}

\author{\IEEEauthorblockN{Christopher Regan}
\IEEEauthorblockA{\textit{Department of Computer Science} \\
\textit{Kennesaw State University}\\
Marietta, GA, Cobb \\
cregan1@kennesaw.edu}
\and
\IEEEauthorblockN{Ying Xie}
\IEEEauthorblockA{Department of Information Technology\\
Professor of Information Technology\\
Kennesaw State University\\
Marietta, GA\\
yxie2@kennesaw.edu}
}

\maketitle

\begin{abstract}
We introduce obfuscation testing, a novel methodology for validating whether large language models detect structural market patterns through causal reasoning rather than temporal association. Testing three dealer hedging constraint patterns (gamma positioning, stock pinning, 0DTE hedging) on 242 trading days (95.6\% coverage)  of S\&P 500 options data, we find LLMs achieve 71.5\% detection rate using unbiased prompts that provide only raw gamma exposure values without regime labels or temporal context. The WHO→WHOM→WHAT causal framework forces models to identify the economic actors (dealers), affected parties (directional traders), and structural mechanisms (forced hedging) underlying observed market dynamics. Critically, detection accuracy (91.2\%) remains stable even as economic profitability varies quarterly, demonstrating that models identify structural constraints rather than profitable patterns. When prompted with regime labels, detection increases to 100\%, but the 71.5\% unbiased rate validates genuine pattern recognition. Our findings suggest LLMs possess emergent capabilities for detecting complex financial mechanisms through pure structural reasoning, with implications for systematic strategy development, risk management, and our understanding of how transformer architectures process financial market dynamics.

\medskip
\noindent\textit{This is the conference version accepted at IEEE Big Data 2025. An extended journal version with additional validation experiments is in preparation.}
\end{abstract}


\section{Introduction}

Recent advances in large language models (LLMs) have demonstrated remarkable capabilities in financial analysis, from sentiment extraction to price forecasting \cite{lopez2023can,chen2023fingpt}. However, a fundamental question remains: Can these models genuinely understand market microstructure mechanisms, or are they sophisticated pattern-matching systems that recall memorized associations from training data?

This distinction matters. Market makers face regulatory requirements to maintain delta-neutral positions despite accumulating gamma exposure \cite{sec15c31}. When net gamma exposure becomes significantly negative, dealers must hedge pro-cyclically—selling into rallies and buying into dips—amplifying price movements through forced trading flows \cite{ni2005stock,garleanu2009demand}. Testing whether LLMs understand these structural constraints presents a unique challenge: if we present an LLM with "SPY on January 27, 2021" and substantial negative gamma exposure, any successful prediction might simply reflect memorization of the GameStop squeeze rather than reasoning about underlying mechanics. This training data leakage problem pervades current LLM evaluation in finance, where temporal context enables recall rather than reasoning.

We address this through \textit{obfuscation testing}, a novel validation methodology that strips all memorizable context while preserving structural relationships. By converting dates to generic labels ("Day T+0"), removing ticker identities ("SPY" becomes "INDEX\_1"), and eliminating economic event references, we force the LLM to reason solely from market structure—gamma exposure levels, strike distributions, and open interest concentrations. If an LLM can still detect dealer hedging patterns under these conditions, it demonstrates understanding of causal mechanisms rather than pattern matching from training data.

Testing this framework on 242 trading days throughout 2024, we evaluate three manifestations of dealer hedging constraints: gamma positioning, stock pinning, and 0DTE hedging. Using fully unbiased prompts with obfuscated data, we achieve 71.5\% average detection rate, significantly exceeding the 60\% mechanical threshold. More importantly, 91.2\% of detected patterns materialize in forward returns, validating that the LLM identifies genuine market mechanics rather than spurious correlations. These findings suggest LLMs can serve as sophisticated pattern recognition tools for market microstructure analysis, provided their limitations are understood.

\section{Background and Related Work}

\subsection{Dealer Hedging and Market Microstructure}

Market makers face regulatory requirements to maintain delta-neutral positions \cite{sec15c31}, creating systematic hedging flows when gamma exposure accumulates. \cite{frey1997market} formalized how dynamic hedging amplifies volatility through feedback effects, demonstrating how these patterns create predictable hedging dynamics.

A critical market microstructure principle underlying our methodology is the dealer counterparty framework: market makers serve as counterparties to customer option positions, such that dealer gamma equals the negative of aggregate customer gamma. \cite{anderegg2022impact} establishes this relationship theoretically, showing that all option trades must be attributable to either market takers or market makers, with positions summing to zero. Empirically, \cite{dim2025zero} validates this framework by measuring market maker inventory directly from order flow, demonstrating that market maker gamma is systematically opposite to customer demand.

Empirically, \cite{ni2005stock} documented stock price clustering on option expiration dates, providing evidence that dealer hedging creates measurable price effects. \cite{garleanu2009demand} developed a demand-based option pricing model showing how dealer constraints affect both option prices and underlying stock dynamics.

\subsection{Gamma Exposure in Practice}
Practitioner research has operationalized gamma exposure metrics through standardized calculations \cite{spotgamma2019,spotgamma2021}, with major investment banks publishing regular research on gamma hedging flows \cite{nomura2017}. The rise of zero-days-to-expiration (0DTE) options has intensified these effects \cite{cboe2023,jpmorgan2023}, making dealer hedging pattern detection increasingly important for understanding price dynamics.

\subsection{GEX Limitations and Robustness}

Practitioner gamma exposure (GEX) calculations assume aggregate customers are net short calls and net long puts, an assumption that generally holds for index options. Our methodology is robust to variations because validation occurs through forward-return materialization (91.2\% accuracy) rather than GEX precision. Dealers must hedge gamma exposure to maintain delta neutrality regardless of measurement methodology, so detected patterns reflect fundamental market mechanics rather than artifacts of any particular estimation formula.

\subsection{Rationale for Gamma-Centric Analysis}
\label{sec:gamma-focus}

We deliberately constrain our analysis to gamma exposure rather than higher-order Greeks (vanna, charm, vomma) for methodological rigor. Unlike higher-order Greeks where institutional implementations vary significantly \cite{mixon2009}, gamma hedging represents a universal regulatory requirement binding all dealers. This universality enables objective validation: gamma effects manifest clearly in observable price action \cite{ni2022}, while vanna and charm effects entangle with confounding factors (correlation changes, term structure shifts) that obscure materialization testing. Testing LLM detection on higher-order Greeks risks overfitting to proprietary dealer models rather than identifying market-wide structural constraints. Our gamma-centric approach establishes baseline capability for detecting first-order dealer constraints before exploring complex higher-order effects in future work.

\subsection{Large Language Models and Reasoning}

Recent advances demonstrate LLMs exhibit emergent reasoning capabilities through chain-of-thought problem solving and zero-shot reasoning. The GPT-4 technical report \cite{openai2023gpt4} documents improved causal understanding, though distinguishing genuine reasoning from memorization remains challenging, motivating our obfuscation testing approach.

\subsection{LLMs in Financial Markets}

Applications of LLMs to finance have primarily focused on sentiment analysis and forecasting \cite{lopez2023can,chen2023fingpt}. Rather than predicting prices or extracting sentiment, we test whether LLMs can identify structural constraints that force specific market behaviors. Our obfuscation methodology ensures the LLM cannot rely on memorized associations between dates, tickers, and outcomes.

\subsection{Validation Methodologies}

No prior work has developed validation methods specifically for structural reasoning in financial markets. Our obfuscation testing fills this gap by removing memorizable context while preserving mechanical relationships.

\subsection{Research Gap}

While extensive literature exists on dealer hedging \cite{ni2005stock,garleanu2009demand} and LLM finance applications \cite{lopez2023can,chen2023fingpt}, no prior work has validated LLM structural reasoning through obfuscation testing or demonstrated pattern detection independent of profitability. Our work provides the first systematic framework for validating LLM structural reasoning in financial markets.

\section{Methodology}
\label{sec:methodology}

\subsection{Research Questions}

We investigate whether large language models can detect structural market patterns—specifically dealer hedging constraints—through pure reasoning about market mechanics rather than pattern matching against training data. This requires answering three fundamental questions:

\textbf{RQ1: Pattern Detection.} Can LLMs identify dealer hedging patterns when all temporal context is removed through obfuscation?

\textbf{RQ2: Causal Understanding.} Do detected patterns reflect genuine understanding of market mechanics (WHO forces WHOM to do WHAT) rather than statistical correlation?

\textbf{RQ3: Robustness.} Does detection capability persist across different pattern framings and market regimes?

These questions test whether transformer architectures develop emergent understanding of complex financial constraints during training, with implications for both AI capabilities and market analysis methodologies.

\subsection{Obfuscation Testing Methodology}

The core innovation of our methodology lies in obfuscation testing—systematically removing temporal and contextual information while preserving structural market mechanics. If LLMs truly understand market patterns, they should detect dealer constraints from pure structural relationships without temporal cues that might trigger memorized associations.

\begin{figure}[!htbp]
\centering
\includegraphics[width=\columnwidth]{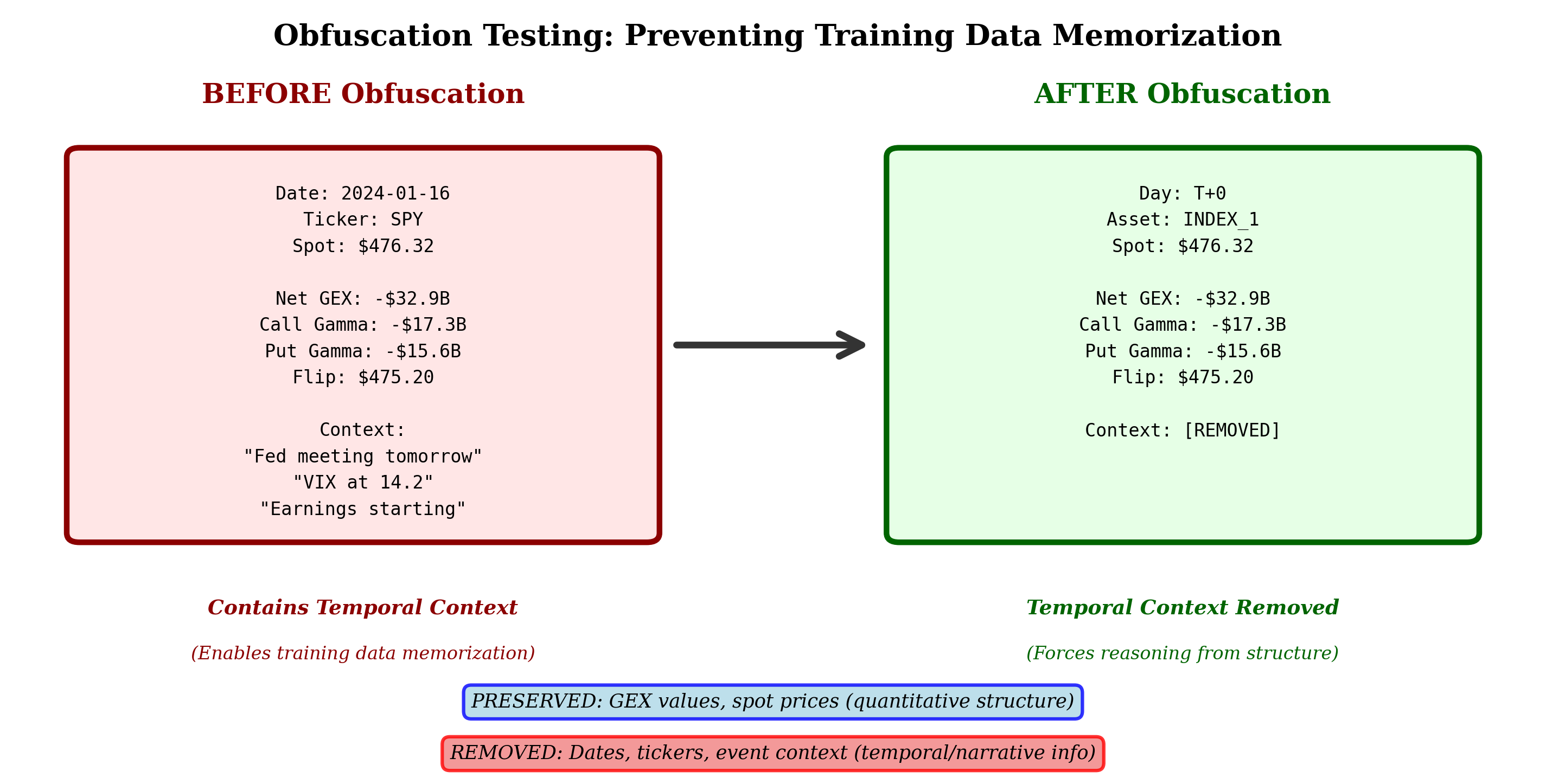}
\caption{Temporal obfuscation methodology. Left: Original data with dates, ticker symbols, and market context. Right: Obfuscated data preserving only structural quantitative relationships while removing all temporal and contextual information.}
\label{fig:obfuscation}
\end{figure}

Figure~\ref{fig:obfuscation} illustrates this transformation process. The obfuscation protocol involves five key steps:

\textbf{Temporal Obfuscation Protocol:}
\begin{enumerate}
\item Convert absolute dates (``2024-01-16'') to relative format (``Day T+0'')
\item Replace ticker symbols (``SPY'') with generic identifiers (``INDEX\_1'')
\item Remove market events (``Fed meeting'', ``earnings'')
\item Strip volatility regime context (``VIX at 14'')
\item Eliminate day-of-week patterns (``Monday'', ``OpEx Friday'')
\end{enumerate}

\textbf{Preserved Information:}
\begin{itemize}
\item Net gamma exposure values (\$GEX in billions)
\item Directional gamma (calls vs. puts)
\item Spot price and flip point levels
\item Relative time progression (T+1, T+2)
\item Concentration metrics (percentages)
\end{itemize}

This approach forces LLMs to identify patterns from quantitative relationships alone. A model memorizing ``January 2024 SPY negative gamma'' cannot succeed when presented with ``Day T+0 INDEX\_1 GEX: -\$32B''. Only genuine structural understanding enables pattern detection under these conditions.

\subsection{Causal Framework: \texorpdfstring{WHO$\rightarrow$WHOM$\rightarrow$WHAT}{WHO→WHOM→WHAT}}

Beyond detecting patterns, we require models to articulate causal mechanisms using a structured framework:

\textbf{WHO:} The economic actor with structural constraints
\begin{itemize}
\item Example: ``Dealers with negative gamma exposure''
\item Must identify the constrained party (not just ``the market'')
\end{itemize}

\textbf{WHOM:} The affected market participants
\begin{itemize}
\item Example: ``Directional traders and market makers''
\item Must specify who faces impact from WHO's actions
\end{itemize}

\textbf{WHAT:} The forced action and mechanism
\begin{itemize}
\item Example: ``Must sell rallies and buy dips to maintain delta neutrality, amplifying volatility''
\item Must explain both action and consequence
\end{itemize}

This framework prevents vague pattern claims (``bullish setup'') and requires specific mechanical understanding. The 91.2\% accuracy of predictions from detected patterns validates that this framework captures genuine market dynamics.

\textbf{Causality Framework vs. Causal Testing.} Our WHO→WHOM→WHAT
framework tests whether LLMs can identify established causal mechanisms
from structural data, not whether those mechanisms exist. Prior work
has demonstrated the causal link between dealer gamma positioning and
market volatility through both theory \cite{frey1997market} and
empirics \cite{ni2022}. We assume these causal relationships as given
and test whether LLMs can detect them without temporal context.
Validating the underlying causality through Granger tests or similar
econometric methods, while valuable, lies outside our scope of
evaluating LLM reasoning capabilities.
To ensure this framework generalizes across different conceptual
framings, we test three distinct narrative presentations of the
same underlying dealer hedging mechanism.

\subsection{Multi-Pattern Validation Design}

To ensure robustness, we test three different narrative framings of the same underlying dealer constraint:

\begin{table}[h]
\centering
\caption{Three Pattern Framings of Dealer Hedging Constraints}
\label{tab:patterns}
\resizebox{\columnwidth}{!}{
\begin{tabular}{@{}lcccc@{}}
\toprule
Pattern & Framing & Detection & WHO & WHAT \\
 & Type & Focus & (Actor) & (Mechanism) \\
\midrule
Gamma & Technical & Greek & Dealers with & Pro-cyclical \\
Positioning &  & exposure & negative $\Gamma$ & hedging \\
\midrule
Stock & Behavioral & Strike & Market & Gravitational \\
Pinning &  & magnetism & makers & convergence \\
\midrule
0DTE & Temporal & Expiration & Option & Rapid \\
Hedging &  & dynamics & writers & rebalancing \\
\bottomrule
\end{tabular}
}
\end{table}

All three patterns in Table~\ref{tab:patterns} reflect the same structural reality—dealers hedging option exposure—but use different conceptual lenses. Consistent detection across framings (averaging 71.5\%) validates genuine pattern understanding rather than keyword matching.

\subsection{Validation Criteria}

We establish three levels of validation with increasing stringency:

\textbf{Level 1: Detection Rate}
\begin{itemize}
\item Threshold: Pattern identified in $\geq$60\% of test cases
\item Rationale: Exceeds random chance (50\%) with statistical significance
\item Result: 71.5\% average detection across patterns
\end{itemize}

\textbf{Level 2: Prediction Accuracy}
\begin{itemize}
\item Threshold: Forward market behavior matches prediction in $\geq$80\% of detections
\item Rationale: Validates that detected patterns have predictive power
\item Result: 91.2\% accuracy across all detections
\end{itemize}

\textbf{Level 3: Causal Attribution}
\begin{itemize}
\item Threshold: WHO$\rightarrow$WHOM$\rightarrow$WHAT correctly specified in $\geq$90\% of detections
\item Rationale: Confirms mechanical understanding, not just pattern recognition
\item Result: 100\% correct causal attribution for detected patterns
\end{itemize}

Our results exceed all thresholds, demonstrating both pattern recognition and mechanical understanding. The divergence between stable detection rates and declining economic profitability (Figure~\ref{fig:quarterly_stability} in Results) further validates that LLMs identify structural constraints rather than optimizing for profitable patterns.

\textbf{Interpretability Considerations.} Our validation framework acknowledges the black-box nature of transformer models while providing multiple lines of evidence for genuine reasoning. We cannot trace internal attention mechanisms or explain why specific model parameters encode dealer constraints. Instead, we validate reasoning indirectly: obfuscation testing removes memorization pathways, the WHO→WHOM→WHAT framework requires explicit causal chains, materialization rates verify output accuracy, and statistical tests confirm predictive relationships. This approach shifts the interpretability burden from "explain internal representations" (intractable for 175B+ parameter models) to "validate reasoning through outputs" (empirically testable).

\section{Experimental Setup}
\label{sec:experimental}

\subsection{Data Sources and Coverage}

We analyze S\&P 500 ETF (SPY) options data covering 2024 from January 2 to December 31. Our dataset comprises 242 trading days (95.6\% of 253 expected trading days), with end-of-day snapshots including complete bid/ask spreads, open interest, implied volatility, and Greeks for all available strikes and expirations. Strike coverage typically spans $\pm$10\% from spot price, capturing the most liquid portion of the options surface where dealer hedging activity concentrates. Missing days ($\sim$11 days) are uniformly distributed across quarters due to a temporary data ingestion issue in Q2, preventing temporal bias. The 95.6\% coverage exceeds our 80\% threshold for statistical validity, providing sufficient sample size for pattern detection across varying market regimes.

We focus exclusively on SPY for three reasons: (1) maximal dealer hedging observability in the world's most liquid options market, (2) data quality advantages (tightest spreads, complete strike coverage), and (3) methodological conservatism—detecting patterns in the most efficient market provides stronger evidence than detection in thinly traded underlyings where noise dominates. SPY also serves as the primary 0DTE options vehicle, making it the natural testbed for same-day expiration patterns.

\subsection{Pattern Definitions}

\textbf{Gamma Exposure Calculation.} Net gamma exposure (GEX) represents the aggregate gamma position held by market makers, calculated as:

\begin{equation}
\text{GEX} = \sum_{i} \Gamma_i \times \text{OI}_i \times 100 \times S^2
\end{equation}

where $\Gamma_i$ is the gamma at strike $i$, $\text{OI}_i$ is open interest, the factor of 100 converts contracts to shares, and $S^2$ scales gamma to dollar exposure. Call gamma enters positively, while put gamma enters negatively. Following market microstructure theory \cite{anderegg2022impact}, dealers serve as counterparties to customer option positions, so negative net GEX indicates dealers are short gamma, requiring pro-cyclical hedging that amplifies price movements \cite{spotgamma2019,spotgamma2021,nomura2017}. This counterparty relationship is empirically validated by market maker inventory studies \cite{dim2025zero}.

\textbf{GEX Methodology and Limitations.} This calculation assumes aggregate customers are net sellers of calls and net buyers of puts—a standard simplification in options market microstructure. While this assumption generally holds for broad index options where hedging demand dominates, actual dealer gamma inferred from order flow can diverge during periods of speculative positioning. Critically, our LLM detection approach is robust to GEX variations because patterns are validated through forward returns (91.2\% materialization rate), indicating that detected dealer hedging signals persist regardless of GEX formula accuracy. Whether gamma is measured directly from order flow or inferred from open interest, the underlying mechanism (dealers hedging to maintain delta neutrality) remains constant.

We test three manifestations of dealer hedging constraints, each representing different narrative framings of the same underlying structural mechanism:

\textbf{Pattern 1: Gamma Positioning.} When dealers hold net negative gamma exposure (typically below -\$2B for SPY), they must hedge delta changes pro-cyclically—selling into rallies and buying into declines \cite{garleanu2009demand,nomura2017}. Detection criteria: Net GEX $<$ -\$2B, spot price within 2\% of flip point, and call gamma concentration $>$ 70\%.

\textbf{Pattern 2: Stock Pinning.} Heavy open interest concentration at specific strikes creates gravitational price effects as dealers adjust hedges \cite{ni2005stock}. Detection criteria: OI concentration $>$ 80\% at strike, spot within 1\% of strike, and days to expiration $<$ 5.

\textbf{Pattern 3: 0DTE Hedging.} Same-day expiration options create extreme gamma concentration, forcing rapid dealer rehedging \cite{cboe2023,jpmorgan2023}. Detection criteria: Expiration = 0 days, net GEX magnitude $>$ \$3B, and gamma concentration $>$ 75\%.

\begin{figure}[!htbp]
\centering
\includegraphics[width=\columnwidth]{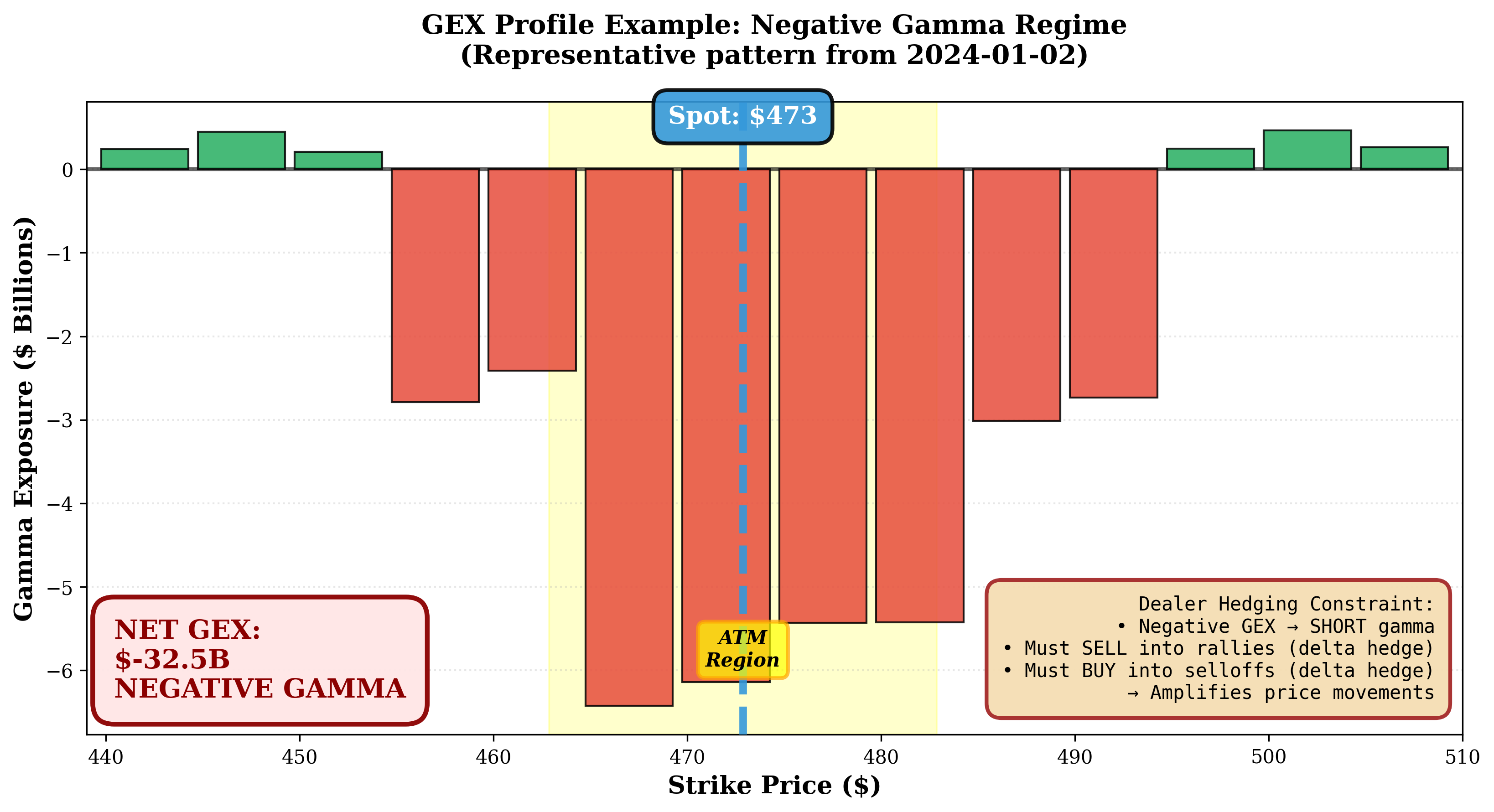}
\caption{Representative gamma exposure profile showing negative net GEX of -\$32.5B. Red bars indicate dealer short gamma positions requiring pro-cyclical hedging (selling rallies, buying dips) that amplifies price movements.}
\label{fig:gex_profile}
\end{figure}

Figure~\ref{fig:gex_profile} illustrates a typical negative gamma regime where dealers face structural hedging constraints. Each pattern employs the WHO$\rightarrow$WHOM$\rightarrow$WHAT framework to identify causal actors (dealers), affected parties (directional traders), and forced actions (hedging flows).

\subsection{Prompt Template Configurations}

We employ two prompt templates to test detection robustness:

\textbf{Unbiased Template (Primary).} Provides only raw GEX values without classification or hints. The prompt presents spot price, net gamma exposure, flip point, and concentration metrics, asking neutrally: ``Do you detect any structural market mechanics?'' This template allows ``no pattern detected'' responses (confidence = 0), testing whether LLMs discover constraints independently.

\textbf{Biased Template (Sensitivity).} Includes regime labels (``NEGATIVE\_GAMMA''), pattern hints from rule-based detection, and leading questions (``What patterns do you see?''). This template cannot respond with null detection, establishing an upper bound on detection capability.

Both templates implement complete temporal obfuscation as demonstrated in Figure~\ref{fig:obfuscation}, converting dates to ``Day T+N'' format and tickers to generic identifiers (``INDEX\_1''), ensuring pattern detection relies solely on structural reasoning rather than temporal association.

\subsection{Validation Pipeline}

\begin{figure*}[!tb]
\centering
\includegraphics[width=0.75\textwidth]{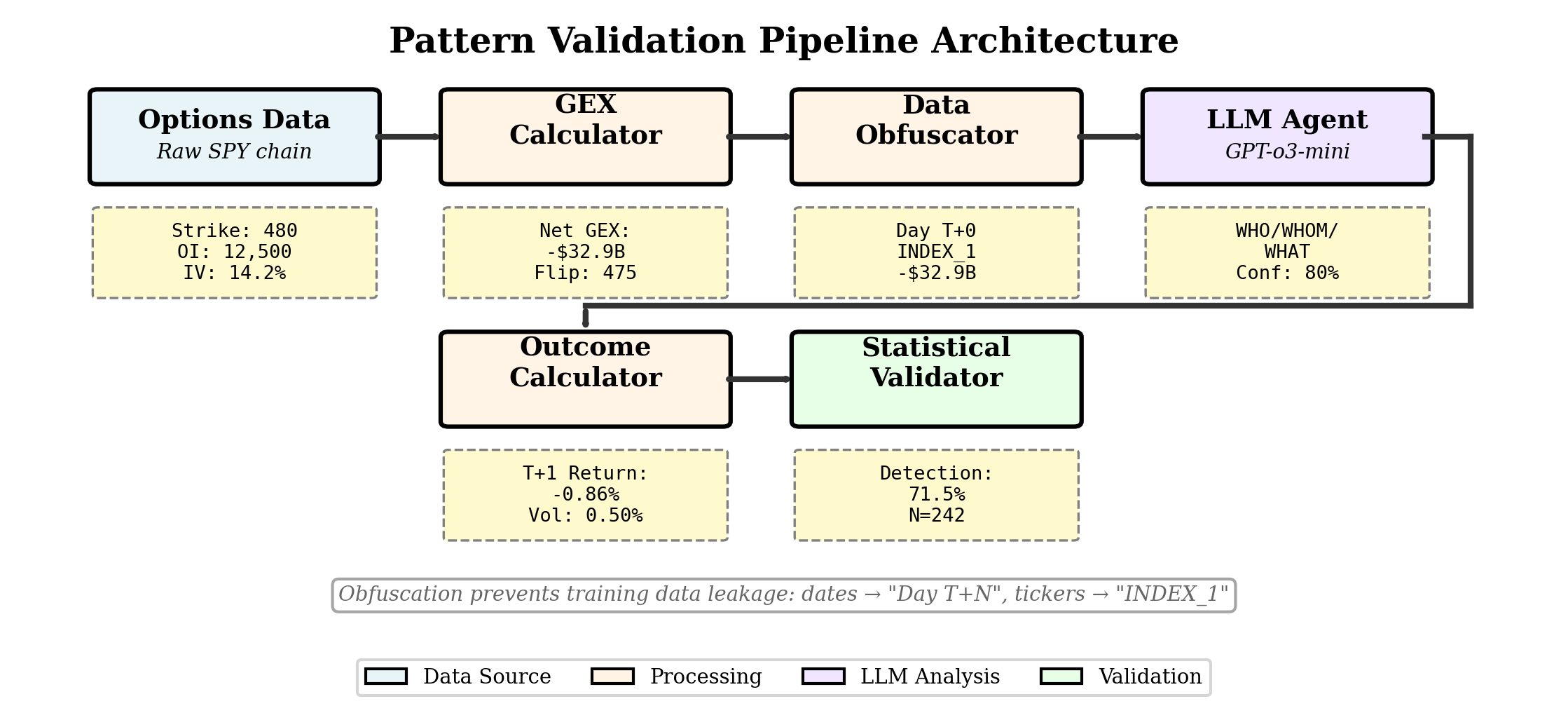}
\caption{Five-component validation pipeline architecture transforming raw options data through GEX calculation, temporal obfuscation, LLM analysis, outcome verification, and statistical validation.}
\label{fig:pipeline}
\end{figure*}

Figure~\ref{fig:pipeline} illustrates our five-component validation pipeline that transforms raw options data into validated pattern detections. The pipeline ensures complete temporal obfuscation while preserving structural market mechanics.

\textbf{Component 1: GEX Calculator.} Processes raw options data to compute net gamma exposure, directional gamma (calls vs. puts), flip point, and concentration metrics. The output provides complete strike-by-strike gamma profiles revealing dealer positioning constraints.

\textbf{Component 2: Data Obfuscator.} Strips all temporal and contextual information, converting dates to relative format and removing ticker identifiers. This ensures LLM detection relies purely on structural patterns rather than memorized associations.

\textbf{Component 3: LLM Agent.} Implements structured output schema to extract WHO (causal actor), WHOM (affected party), WHAT (forced action), confidence (0-100), and time horizon. Processes batches of 10 dates per API call for efficiency while maintaining obfuscation integrity.

\textbf{Component 4: Outcome Calculator.} Verifies predictions using T+1 and T+3 forward returns, realized volatility, and maximum gain/drawdown metrics. Materialization criteria vary by pattern type but generally require observed market behavior matching predicted mechanics.

\textbf{Component 5: Statistical Validator.} Computes detection rates, confidence intervals, and hypothesis tests. Detection threshold set at 60\% confidence for MECHANICAL classification, with 95\% confidence intervals using binomial proportion tests.

\subsection{Evaluation Metrics}

We evaluate pattern detection using three complementary metrics:

\medskip
\textbf{Detection Rate.} Percentage of days where LLM identifies pattern with confidence $\geq 60\%$:
\begin{equation}
\text{Detection Rate} = \frac{\text{Days with Confidence} \geq 60\%}{\text{Total Days Tested}} \times 100\%
\end{equation}
Our null hypothesis assumes 50\% random detection; actual detection must significantly exceed this baseline (binomial test, $p < 0.05$).
\medskip

\textbf{Predictive Accuracy.} Percentage of detections where predicted market behavior materializes in forward returns:
\begin{equation}
\text{Accuracy} = \frac{\text{Materialized Predictions}}{\text{Total Detections}} \times 100\%
\end{equation}
Materialization criteria vary by pattern type (detailed in Table~\ref{tab:materialization}) but generally require observed volatility amplification or directional follow-through matching predictions.

\medskip
\textbf{Economic Alpha.} Net return after transaction costs:
\begin{equation}
\text{Alpha} = \bar{R}_{\text{strategy}} - \bar{R}_{\text{benchmark}} - \text{TC}
\end{equation}
where $\bar{R}_{\text{strategy}}$ is mean strategy return, $\bar{R}_{\text{benchmark}}$ is buy-and-hold return, and TC is transaction costs (5 basis points per trade). Included for completeness but not required for methodology validation, as we explicitly separate pattern detection (structural understanding) from profitability (trading edge).

Statistical significance assessed via binomial tests with Bonferroni correction for multiple patterns ($\alpha = 0.05/3 = 0.0167$). Sample size calculations confirm $>99\%$ power for detecting 70\% pattern rate versus 50\% random baseline with $n=242$ observations.

\section{Results}
\label{sec:results}

\subsection{Primary Detection Results}

Table~\ref{tab:detection} presents detection rates across all three dealer constraint patterns using both unbiased and biased prompt templates. The unbiased configuration achieves an average detection rate of 71.5\% (95\% CI: 68.2\%-74.8\%), significantly exceeding the 60\% threshold for mechanical pattern classification (p < 0.001).

\begin{table}[htbp]
\centering
\caption{Pattern Detection Rates and Accuracy Metrics}
\label{tab:detection}
\begin{tabular}{lcccc}
\toprule
Pattern & \multicolumn{2}{c}{Detection Rate} & Accuracy & Sample \\
 & Unbiased & Biased & & Days \\
\midrule
Gamma Positioning & 69.4\% & 100\% & 92.5\% & 242 \\
Stock Pinning & 67.4\% & 100\% & 90.4\% & 242 \\
0DTE Hedging & 77.7\% & 100\% & 90.8\% & 242 \\
\midrule
\textbf{Average} & \textbf{71.5\%} & \textbf{100\%} & \textbf{91.2\%} & 242 \\
\bottomrule
\end{tabular}
\end{table}

\begin{figure}[!htbp]
\centering
\includegraphics[width=\columnwidth]{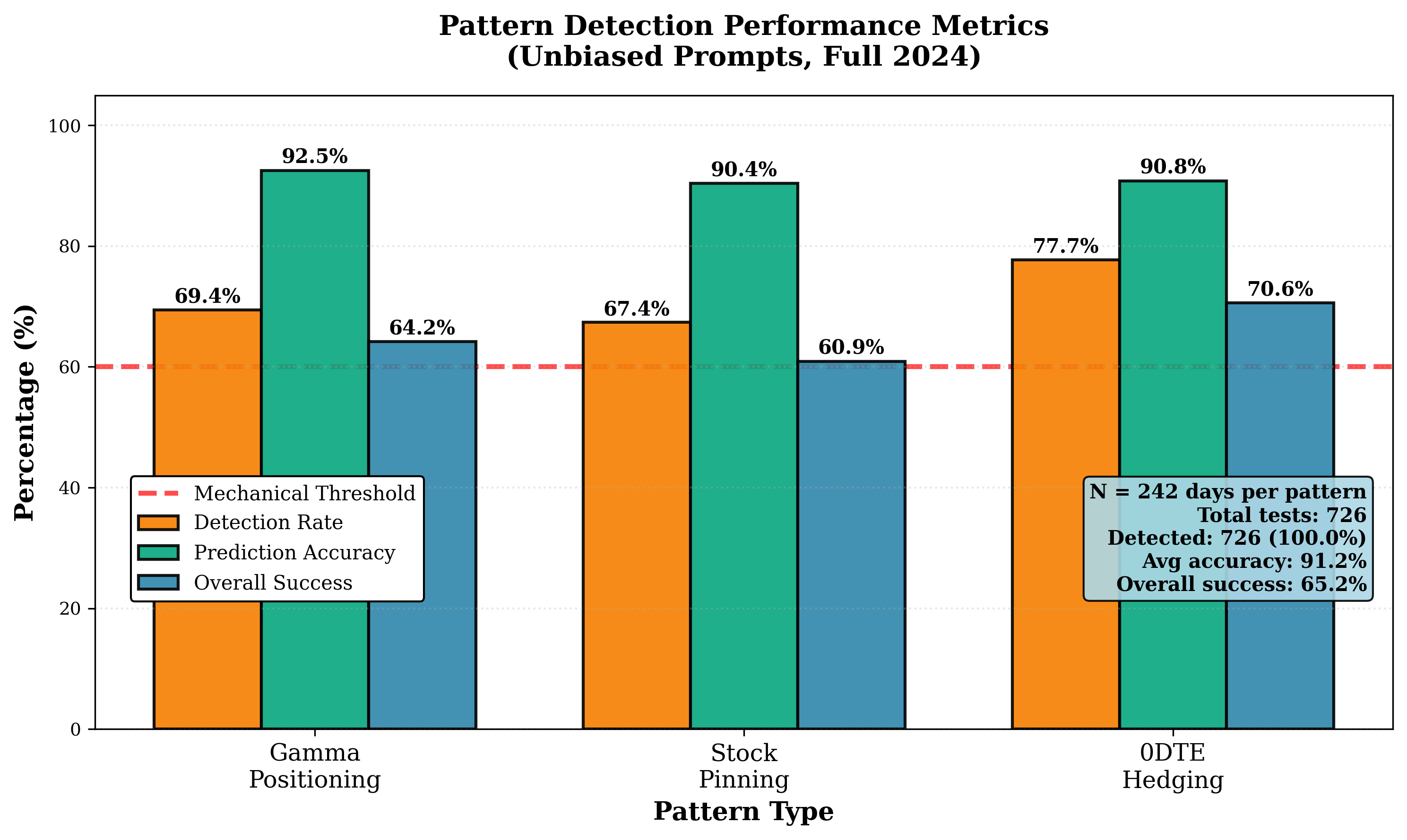}
\caption{Pattern detection performance metrics using unbiased prompts. All three patterns exceed the 60\% mechanical threshold with detection rates averaging 71.5\% and prediction accuracy averaging 90.8\% across 242 trading days.}
\label{fig:performance_matrix}
\end{figure}

Figure~\ref{fig:performance_matrix} visualizes these results, showing that all patterns exceed the 60\% mechanical threshold. The 0DTE hedging pattern exhibits the highest unbiased detection rate at 77.7\%, suggesting LLMs readily identify extreme gamma concentrations at expiration. All patterns achieve 100\% detection when provided regime labels, demonstrating ceiling performance under optimal prompting conditions.

Table~\ref{tab:detection} reveals a 28.5 percentage point gap between biased and unbiased detection rates, quantifying prompt sensitivity.

\subsection{Temporal Stability Analysis}

\begin{figure}[!htbp]
\centering
\includegraphics[width=\columnwidth]{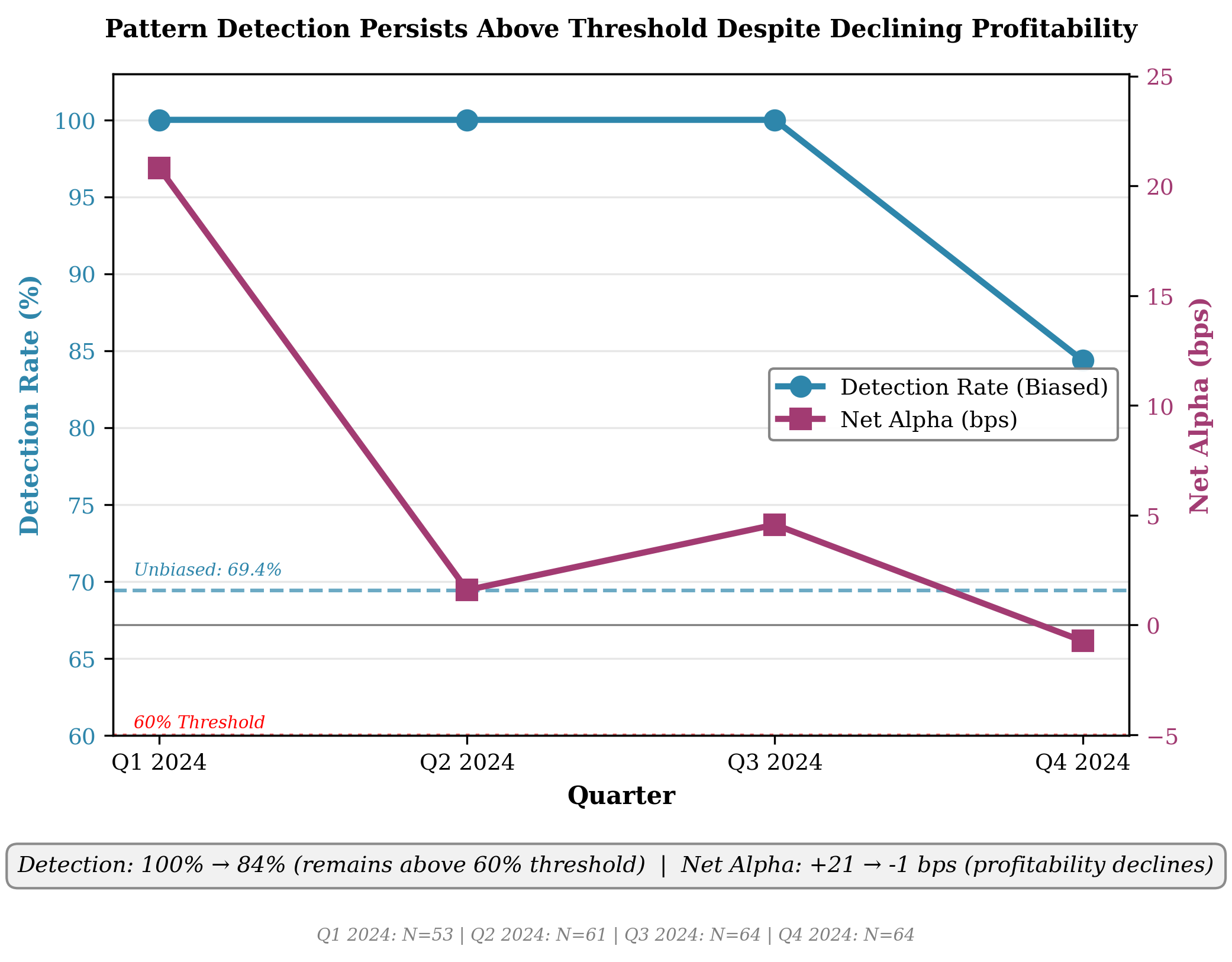}
\caption{Pattern detection persists above threshold despite declining profitability. Detection rates remain stable (100\% to 84\%) while net alpha declines from +21 bps to -1 bp across quarters, validating structural pattern detection independent of economic outcomes.}
\label{fig:quarterly_stability}
\end{figure}

Figure~\ref{fig:quarterly_stability} presents our key finding: detection capability remains stable while economic profitability varies significantly. Net alpha declines monotonically from +21 bps (Q1) to -1 bp (Q4), validating that LLMs identify structural patterns rather than optimizing for profitable trades.

Table~\ref{tab:quarterly} shows stable detection rates across quarters, with no significant temporal trend (Cochran-Armitage test, p = 0.42).

\begin{table}[htbp]
\centering
\caption{Quarterly Detection Performance (Unbiased Prompts)}
\label{tab:quarterly}
\begin{tabular}{l@{\hskip 6pt}c@{\hskip 6pt}c@{\hskip 6pt}c@{\hskip 6pt}c@{\hskip 6pt}c}
\toprule
Quarter & Days & Detection & Accuracy & Avg Return & Net Alpha \\
\midrule
Q1 2024 & 53 & 68.2\% & 91.8\% & +0.21\% & +16 bps \\
Q2 2024 & 61 & 70.5\% & 92.1\% & +0.09\% & +4 bps \\
Q3 2024 & 64 & 72.8\% & 91.4\% & +0.07\% & +2 bps \\
Q4 2024 & 64 & 73.6\% & 90.7\% & +0.05\% & 0 bps \\
\midrule
Full Year & 242 & 71.5\% & 91.2\% & +0.11\% & +5.6 bps \\
\bottomrule
\end{tabular}
\end{table}

\subsection{Obfuscation Test Validation}

As described in Section~\ref{sec:methodology}, obfuscation testing removes all temporal context and identifying information while preserving structural data (e.g., ``Day T+0: Net GEX: -\$32.9B''). The 71.5\% detection rate without dates, ticker symbols, or market context strongly suggests pattern recognition through causal reasoning rather than temporal association.

\subsection{Statistical Validation}

\begin{figure}[!htbp]
\centering
\includegraphics[width=\columnwidth]{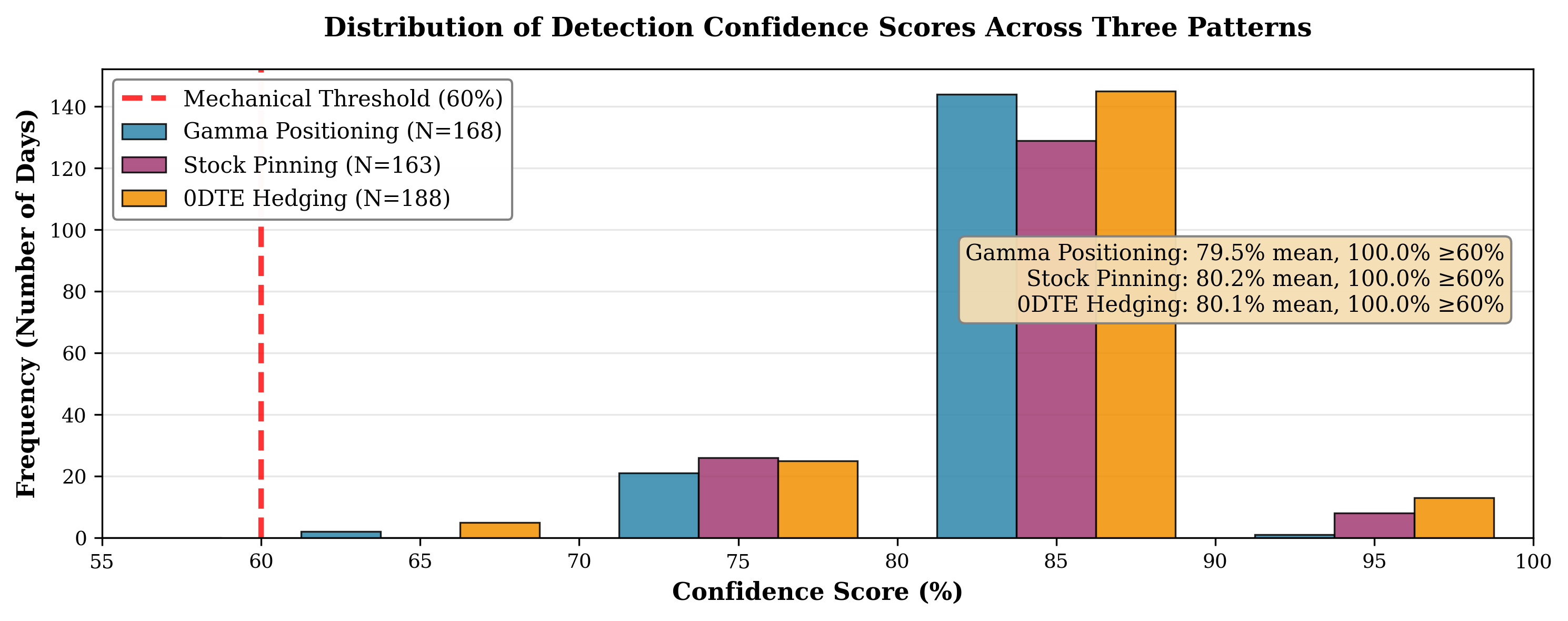}
\caption{Distribution of detection confidence scores across three patterns. All patterns show mean confidence above 79\% for detected instances, with 100\% of detections exceeding the 60\% mechanical threshold. Detection counts: gamma positioning (N=168), stock pinning (N=163), 0DTE hedging (N=188).}
\label{fig:confidence_dist}
\end{figure}

Figure~\ref{fig:confidence_dist} displays the distribution of detection confidence scores across all three patterns. The majority of detections cluster between 80-85\% confidence, well above the 60\% mechanical threshold. Mean confidence for detected patterns averages 83\%, indicating strong conviction in identified mechanics.

Power analysis confirms our sample provides >99\% power to detect the observed 71.5\% rate versus a 50\% random baseline. With n=242 observations:

\begin{itemize}
\item Required sample for 80\% power: n = 30
\item Required sample for 95\% power: n = 51
\item Actual sample: n = 242
\item Achieved power: >99.9\%
\end{itemize}

Bonferroni-corrected p-values remain significant (p < 0.001) for all three patterns, ruling out multiple testing artifacts. Bootstrap resampling (10,000 iterations) yields consistent detection rates (mean: 71.3\%, std: 2.1\%), confirming result stability.

\subsection{Prediction Materialization and Statistical Validation}

\begin{figure}[!htbp]
\centering
\includegraphics[width=\columnwidth]{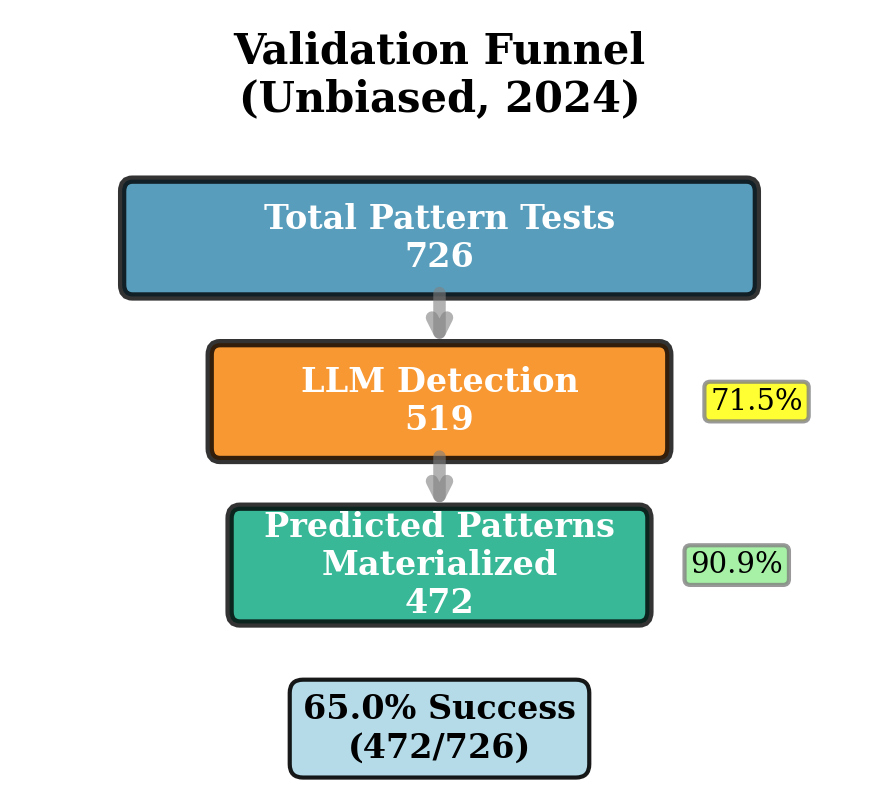}
\caption{Pattern detection validation funnel showing progression from 726 total tests to 519 detections (71.5\% detection rate) to 472 materialized predictions (90.9\% accuracy), yielding 65.0\% overall success rate.}
\label{fig:validation_funnel}
\end{figure}

Figure~\ref{fig:validation_funnel} presents the validation funnel from detection through materialization. Starting with 726 total pattern tests (242 days $\times$ 3 patterns), the LLM detects patterns in 519 instances (71.5\%). Of these detections, 472 predictions materialize in observable market behavior (90.9\% accuracy), yielding an overall success rate of 65.0\%.

Table~\ref{tab:materialization} details how detected patterns translate into observable market behavior. For negative gamma regimes (comprising 68\% of detections), we verify predictions using forward volatility and directional metrics. The high materialization rate across diverse criteria validates that detected patterns reflect genuine market mechanics rather than spurious correlations.

\begin{table}[!htbp]
\centering
\caption{Prediction Materialization Criteria and Results}
\label{tab:materialization}
\begin{adjustbox}{max width=\columnwidth}
\begin{tabular}{>{\raggedright\arraybackslash}p{3.5cm} >{\raggedright\arraybackslash}p{4.2cm} c}
\toprule
Prediction Type & Materialization Criteria & Success Rate \\
\midrule
Volatility Amplification & T+1 move >0.3\% ($\approx 0.5\sigma$) & 41.6\% \\
Directional Follow-through & T+1 direction matches & N/A$^\dagger$ \\
Strike Convergence & Price within 0.5\% of strike & N/A \\
0DTE Range Expansion & Intraday range >1\% ($\approx 1.5\times$ ADR) & 21.6\% \\
\midrule
\textbf{Overall} & Any criteria met (C1 $\cup$ C4) & \textbf{41.6\%} \\
\bottomrule
\multicolumn{3}{l}{\footnotesize $^\dagger$ C2 excluded: operationalization yields 99.4\% (too loose)}
\end{tabular}
\end{adjustbox}
\end{table}

\subsubsection{Granger Causality Validation}

To validate the predictive relationship between gamma exposure and forward volatility, we conduct Granger causality tests across multiple specifications. Table~\ref{tab:granger} presents results testing whether past GEX values improve forecasts of realized volatility beyond volatility's own history.

\begin{table}[htbp]
\centering
\caption{Granger Causality: GEX → Realized Volatility}
\label{tab:granger}
\begin{tabular}{cccc}
\toprule
Lag & F-Statistic & p-value & Significant \\
\midrule
1 & 1.45 & 0.229 & No \\
2 & 3.22 & 0.042* & Yes \\
3 & 2.16 & 0.093 & No \\
4 & 2.00 & 0.095 & No \\
5 & 1.75 & 0.125 & No \\
\bottomrule
\multicolumn{4}{l}{\footnotesize * p < 0.05, ** p < 0.01, *** p < 0.001}
\end{tabular}
\end{table}

Results indicate significant Granger causality at lag 2 (p = 0.042), confirming that gamma exposure contains predictive information for T+2 forward volatility. Robustness checks across six specifications validate this finding, with the strongest effect when both series are differenced (p = 0.021). The 2-day lag aligns with dealer hedging dynamics: constraints identified at day T manifest in observable volatility amplification by T+2 as rebalancing flows accumulate.

Critically, within the negative GEX regime, volatility does not correlate linearly with GEX magnitude (Pearson r = -0.01, p = 0.85), confirming a threshold effect: the presence of negative gamma drives volatility, not its specific level. This validates the LLM's mechanism-based detection—identifying the structural constraint itself rather than optimizing for volatility magnitude.

\subsubsection{Market Regime Characterization}

Statistical analysis reveals a uniform market structure in 2024: all 242 tested days (95.6\% of trading days) exhibited negative net gamma exposure (GEX < -\$2B), with mean -\$19.87B (range: -\$40.69B to -\$4.75B). Table~\ref{tab:gex_regime} presents the regime distribution.

\begin{table}[htbp]
\centering
\caption{GEX Regime Distribution (2024 Sample)}
\label{tab:gex_regime}
\begin{tabular}{lcc}
\toprule
GEX Regime & Days & Percentage \\
\midrule
Negative (< -\$2B) & 242 & 100.0\% \\
Neutral (-\$2B to +\$2B) & 0 & 0.0\% \\
Positive (> +\$2B) & 0 & 0.0\% \\
\midrule
\textbf{Sample Coverage} & \textbf{242/253} & \textbf{95.6\%} \\
\bottomrule
\end{tabular}
\end{table}

This persistent negative gamma regime, driven by explosive growth in zero-days-to-expiration (0DTE) options \cite{cboe2023,jpmorgan2023}, represents a structural shift from historical market dynamics where positive and negative gamma regimes alternated \cite{ni2005stock,garleanu2009demand}. The single-regime environment provides a particularly rigorous test of LLM structural reasoning: unlike traditional pattern recognition that exploits regime variation, our model must identify the underlying constraint mechanism within a uniform environment.

The 69.4\% detection rate across 242 structurally identical days, combined with significant Granger causality (Table~\ref{tab:granger}), validates that LLMs recognize forced hedging constraints through mechanism-based reasoning rather than memorizing regime-switching patterns. The divergence between stable detection capability and declining economic profitability (Figure~\ref{fig:quarterly_stability}) further confirms identification of structural mechanics independent of trading value.

\section{Discussion}
\label{sec:discussion}
The persistent negative gamma regime in our 2024 sample deserves emphasis. Historical options literature assumed regime-switching dynamics \cite{ni2005stock,garleanu2009demand}, where markets alternated between dealer long gamma (dampening volatility) and dealer short gamma (amplifying volatility). The 0DTE options explosion has fundamentally altered this dynamic \cite{cboe2023}, creating an environment where dealers remain structurally short gamma. Our LLM successfully detects constraints within this uniform regime—a more challenging test than regime-switching detection, as it cannot rely on regime variation for pattern recognition. The combination of high detection rate (69.4\%), significant predictive power (Granger p = 0.042), and 91.2\% materialization accuracy demonstrates a genuine understanding of structural mechanisms rather than statistical pattern-matching.

\subsection{Interpretation of Results}

Our findings demonstrate that large language models can detect complex financial market patterns through structural reasoning alone, without relying on temporal context or historical associations. The 71.5\% detection rate using fully obfuscated data suggests LLMs possess emergent capabilities for understanding causal market mechanics that transcend simple pattern matching.

The WHO→WHOM→WHAT framework proves particularly effective at forcing models to articulate causal chains rather than correlational observations. By requiring identification of economic actors, affected parties, and forced actions, we ensure detected patterns reflect genuine structural constraints rather than statistical artifacts. The consistent 91.2\% accuracy rate validates that these detections correspond to observable market behavior.

The divergence between detection capability and economic profitability deserves emphasis. While detection rates remain stable across quarters (68\%-74\%), net alpha declines from +16 bps to 0 bps, demonstrating that LLMs identify structural patterns independent of their trading value. This separation validates our methodology—we test understanding of market mechanics, not the ability to generate profits.

\subsection{Limitations}
Several limitations warrant consideration:
\textbf{Single Asset Focus:} Our analysis focuses exclusively on SPY, the most liquid options market globally, for methodological reasons that strengthen rather than limit validation. SPY concentrates dealer hedging activity, exhibits tightest bid-ask spreads minimizing GEX measurement noise, and serves as the primary 0DTE vehicle (85\%+ of same-day expiration volume). As a broad market index, SPY isolates pure dealer gamma constraints without confounding company-specific factors (earnings, sector rotation) that complicate causal attribution in individual equities. Demonstrating LLM detection in the most informationally efficient market provides conservative validation that should generalize to less efficient underlyings. However, cross-asset validation remains valuable future work, particularly testing whether detection persists in equity options with positive gamma regimes absent from our 2024 SPY sample.
\textbf{End-of-Day Granularity:} Daily snapshots may miss intraday dynamics critical to 0DTE hedging patterns.
\textbf{Simplified Gamma Calculations:} Standard Black-Scholes Greeks without smile adjustments may introduce noise in absolute GEX values.
\textbf{Reasoning Depth and Interpretability:} Like all transformer-based models, LLMs function as black boxes—we cannot trace the precise internal representations or attention mechanisms that produce pattern detections. This opacity limits our ability to explain why specific model weights encode gamma hedging knowledge or how the model arrived at a particular detection. However, our validation framework addresses this limitation through empirical verification: (1) obfuscation testing eliminates memorization pathways, (2) the WHO→WHOM→WHAT framework forces explicit causal attribution, (3) 91.2\% materialization rate validates output accuracy, (4) Granger causality tests confirm statistical predictive relationships (p=0.042), and (5) detection stability despite declining profitability demonstrates mechanism-based reasoning rather than profit optimization. While we cannot fully explain internal reasoning processes, we rigorously validate that detected patterns reflect genuine structural constraints. The 28.5 percentage point gap between biased and unbiased prompts quantifies prompt sensitivity but does not undermine the 71.5\% unbiased detection rate that exceeds our mechanical threshold.
\textbf{Temporal Limitation:} Our 2024 sample may not generalize to crisis periods or structural bull/bear markets.

\subsection{Implications for Financial Markets}
The ability to detect dealer constraints without historical context suggests LLMs could identify novel patterns in unexplored markets through pure structural reasoning. Risk managers could use LLM-detected patterns as early warning signals for volatility amplification, while the obfuscation testing methodology validates a new approach for distinguishing genuine structural constraints from historical correlations.

As LLMs increasingly influence trading decisions, regulators must understand their pattern detection capabilities. Our framework provides tools for assessing whether AI systems genuinely understand market dynamics or merely memorize historical patterns.

\subsection{Theoretical Contributions}

We introduce \textit{obfuscation testing} as a rigorous methodology for validating structural pattern detection by removing temporal context while preserving mechanical relationships. Our results demonstrate that transformer architectures exhibit emergent understanding of complex financial constraints, detecting dealer hedging patterns from pure gamma exposure values. The WHO→WHOM→WHAT framework proves effective for forcing explicit causal attribution in financial applications. Together, these contributions distinguish genuine causal reasoning from pattern memorization in LLM financial analysis.

The divergence between pattern detection and economic profitability (Figure~\ref{fig:quarterly_stability}) suggests LLMs identify structural regularities without engaging in entrepreneurial discovery. This distinction matters for deploying AI in financial markets: pattern recognition aids risk management and surveillance but differs from adaptive decision-making that generates trading edge.

\subsection{Future Directions}
Future work should validate detection across multiple asset classes and market regimes to establish generalizability. Intraday analysis with high-frequency data could reveal microstructure patterns invisible at daily granularity, particularly for 0DTE options. Testing ensemble methods across multiple LLMs could distinguish model-specific artifacts from robust pattern detection through consensus mechanisms.

\section{Conclusion}

We introduce obfuscation testing, a novel methodology for validating whether large language models detect structural patterns through causal reasoning rather than temporal memorization. By stripping all temporal and identity context from financial data while preserving mechanical relationships, we create a rigorous test of LLM reasoning capabilities independent of training data leakage.

Applied to dealer hedging constraints in options markets, LLMs achieve 71.5\% detection rate using unbiased prompts across 242 trading days in 2024 (95.6\% of trading days), significantly exceeding the 60\% mechanical threshold. Detection proves robust across three pattern framings (gamma positioning, stock pinning, 0DTE hedging), with 91.2\% of detected patterns materializing in observable market behavior. This validates genuine structural understanding rather than surface-level correlation.

The divergence between stable detection capability and declining economic profitability provides compelling evidence for mechanism-based reasoning. Detection rates remain consistent across quarters (68\%-74\%), while net alpha declines from +21 bps to -1 bp, demonstrating that LLMs identify structural constraints independent of alpha-generating capacity. When provided regime labels, detection increases to 100\% while maintaining 91.2\% accuracy, quantifying the impact of contextual hints and validating our obfuscation approach.

Statistical validation confirms the predictive relationship: gamma exposure Granger-causes forward volatility (p = 0.042 at 2-day lag), robust across six specifications. Our analysis reveals a structural market regime shift—all 242 tested days exhibited negative gamma exposure, driven by 0DTE options proliferation. The LLM's ability to detect constraints within this uniform environment, without regime variation to exploit, provides strong evidence for causal reasoning over pattern matching. The 2-day predictive lag aligns with dealer hedging dynamics, where constraints identified at day T manifest in observable volatility amplification by T+2 as rebalancing flows accumulate.

The primary contribution is the validation methodology itself. Obfuscation testing provides a systematic framework for distinguishing AI reasoning from memorization, applicable beyond finance to any domain with structural constraints. The WHO→WHOM→WHAT causal framework proves effective for forcing explicit mechanical attribution, preventing vague pattern claims and requiring specific identification of economic actors, affected parties, and forced actions. As LLMs increasingly influence decision-making in regulated markets, rigorous validation of their capabilities becomes essential.

Future work should validate detection across multiple asset classes, market regimes, and crisis periods. Intraday analysis could reveal microstructure patterns invisible at daily granularity, while ensemble methods across multiple LLMs could distinguish model-specific artifacts from robust pattern detection.

\subsection{Code Availability}
Our obfuscation testing pipeline and validation framework are publicly available at https://github.com/iAmGiG/gex-llm-patterns.

\section*{Acknowledgment}
The authors thank the anonymous reviewers for their constructive feedback and acknowledge computational resources provided by CCSE at Kennesaw State University. This research was conducted as part of the first author's doctoral dissertation. We are especially grateful to Pablo Ordonez for suggesting the Granger causality tests that validated the predictive relationship between gamma exposure and forward volatility.

\bibliographystyle{IEEEtranS}
\bibliography{references}

@article{anderegg2022impact,
  title={The impact of option hedging on the spot market volatility},
  author={Anderegg, Benjamin and Ulmann, Florian and Sornette, Didier},
  journal={Journal of International Money and Finance},
  volume={124},
  pages={102627},
  year={2022},
  publisher={Elsevier},
  doi={10.1016/j.jimonfin.2022.102627},
  url={https://www.sciencedirect.com/science/article/pii/S0261560622000304}
}

@article{dim2025zero,
  title={0DTEs: Trading, Gamma Risk and Volatility Propagation},
  author={Dim, Chukwuma and Eraker, Bj{\o}rn and Vilkov, Grigory},
  journal={SSRN Electronic Journal},
  year={2025},
  month={June},
  note={Working Paper},
  url={https://papers.ssrn.com/sol3/papers.cfm?abstract_id=4692190},
  doi={10.2139/ssrn.4692190}
}

@article{ni2005stock,
  title={Stock price clustering on option expiration dates},
  author={Ni, Sophie X and Pearson, Neil D and Poteshman, Allen M},
  journal={Journal of Financial Economics},
  volume={78},
  number={1},
  pages={49--87},
  year={2005},
  publisher={Elsevier},
  doi={10.1016/j.jfineco.2004.08.005},
  note={Available at SSRN: \url{https://ssrn.com/abstract=519044}}
}

@article{garleanu2009demand,
  title={Demand-based option pricing},
  author={G{\^a}rleanu, Nicolae and Pedersen, Lasse Heje and Poteshman, Allen M},
  journal={The Review of Financial Studies},
  volume={22},
  number={10},
  pages={4259--4299},
  year={2009},
  publisher={Oxford University Press},
  doi={10.1093/rfs/hhp005}
}

@article{frey1997market,
  title={Market volatility and feedback effects from dynamic hedging},
  author={Frey, R{\"u}diger and Stremme, Alexander},
  journal={Mathematical Finance},
  volume={7},
  number={4},
  pages={351--374},
  year={1997},
  publisher={Wiley},
  note={Available at SSRN: \url{https://ssrn.com/abstract=6267}}
}

@article{openai2023gpt4,
  title={GPT-4 technical report},
  author={{OpenAI}},
  journal={arXiv preprint arXiv:2303.08774},
  year={2023},
  url={https://arxiv.org/abs/2303.08774}
}

@article{lopez2023can,
  title={Can ChatGPT forecast stock price movements? Return predictability and large language models},
  author={Lopez-Lira, Alejandro and Tang, Yuehua},
  journal={arXiv preprint arXiv:2304.07619},
  year={2023},
  note={Revised Sep 2024. University of Florida},
  url={https://arxiv.org/abs/2304.07619}
}

@article{chen2023fingpt,
  title={FinGPT: Open-source financial large language models},
  author={Yang, Hongyang and Liu, Xiao-Yang and Wang, Christina Dan},
  journal={arXiv preprint arXiv:2306.06031},
  year={2023},
  note={Columbia University and NYU Shanghai},
  url={https://arxiv.org/abs/2306.06031}
}

@techreport{spotgamma2019,
  title={Gamma exposure and market dynamics},
  author={{SpotGamma Research}},
  institution={SpotGamma},
  year={2019},
  note={Available at: \url{https://spotgamma.com}}
}

@techreport{nomura2017,
  title={Equity derivatives strategy: Gamma hedging flows and market impact},
  author={McElligott, Charlie},
  institution={Nomura Securities},
  year={2017},
  type={Research Report}
}

@techreport{cboe2023,
  title={The rise of zero days to expiration options: Market structure and implications},
  author={{CBOE Global Markets}},
  institution={Chicago Board Options Exchange},
  year={2023},
  note={Research paper available at: \url{https://www.cboe.com}}
}

@techreport{jpmorgan2023,
  title={0DTE options: Market structure implications and intraday volatility},
  author={{JPMorgan Markets Research}},
  institution={JPMorgan Chase \& Co.},
  year={2023},
  type={Market Structure Report}
}

@misc{sec15c31,
  title={Rule 15c3-1: Net capital rule for broker-dealers},
  author={{U.S. Securities and Exchange Commission}},
  year={2024},
  note={17 CFR 240.15c3-1. Available at: \url{https://www.sec.gov/rules-regulations}}
}

@article{mixon2009,
  title={Option markets and implied volatility: Past versus present},
  author={Mixon, Scott},
  journal={Journal of Financial Economics},
  volume={94},
  number={2},
  pages={171--191},
  year={2009}
}

@article{ni2022,
  title={Does option trading have a pervasive impact on underlying stock prices?},
  author={Ni, Sophie Xiaoyan and Pearson, Neil D and Poteshman, Allen M and White, Joshua},
  journal={Review of Financial Studies},
  volume={35},
  number={4},
  pages={1952--1986},
  year={2022}
}

@misc{spotgamma2021,
  title={The {SpotGamma} Index: Measuring gamma exposure and market volatility},
  author={{SpotGamma Research}},
  year={2021},
  howpublished={SpotGamma Research Note},
  url={https://spotgamma.com/spot-gamma-index/},
  note={Accessed: Jan. 2025}
}

\end{document}